# Thermodynamics of pressure activated disassembly of supramolecules in isochoric and isobaric systems.

Boris Rubinsky
Department of Mechanical Engineering
University of California Berkeley, Berkeley CA 94720 USA
rubinsky@berkeley.edu

**Abstract**

The efficacy of cryopreservation is fundamentally limited by the difficulty of achieving sufficiently high intracellular concentrations of cryoprotective solutes without inducing osmotic injury or chemical toxicity during loading. This work presents a thermodynamic hypothesis, rather than an experimental demonstration, in which effective cryoprotection could be achieved through the *in situ* generation of cryoprotective solutes via pressure-activated disassembly of supramolecular assemblies composed of cryoprotectant monomers or oligomers. We propose that elevated hydrostatic pressure, whether generated intrinsically during isochoric (constant-volume) freezing or applied externally under isobaric conditions, can destabilize supramolecular assemblies whose dissociated state occupies a smaller molar volume than the assembled state. Under isochoric freezing, ice formation within a fixed volume produces a substantial increase in pressure as a thermodynamic consequence of phase change, rendering pressure a dependent variable governed by the Helmholtz free energy. In contrast, under isobaric conditions, pressure enters explicitly as a control variable through the Gibbs free energy. In both formulations, increasing pressure reshapes the free-energy landscape in a manner that can favor disassembly when the assembled state has a positive excess molar volume. If operative, pressure-activated disassembly would decouple membrane transport from cryoprotectant availability and enable synchronized solute generation precisely during cooling or freezing, without reliance on biochemical triggers or pre-loading of osmotically active solutes. While isochoric freezing offers a natural and self-synchronized mechanism for generating pressure during cryopreservation, the underlying thermodynamic principle is general and applies equally to externally pressurized systems. The purpose of this contribution is to articulate a physically plausible thermodynamic framework that identifies pressure, regardless of its mode of generation, as an underutilized control variable in cryopreservation, and to motivate future experimental and theoretical investigations of pressure-mediated preservation strategies for cells, tissues, and organs.

Keywords: Thermodynamic hypothesis, cryopreservation; isochoric freezing; Helmholtz free energy; hydrostatic pressure; supramolecular assembly; nanoparticles

---

**Introduction**

Intracellular ice formation is widely recognized as a dominant cause of cryoinjury in cryopreserved biological systems. Effective suppression of ice nucleation and growth requires sufficiently high intracellular concentrations of cryoprotective solutes. Conventional cryopreservation strategies therefore rely on transporting these solutes across cell membranes prior to cooling, tightly coupling cryoprotection to osmotic stress, chemical toxicity, and long equilibration times. These constraints are particularly severe in cells with low membrane permeability and in multicellular tissues and organs, where diffusion distances further limit solute delivery. These well-established limitations motivate the exploration of alternative conceptual frameworks for intracellular cryoprotection.

In this paper, we advance a hypothesis, rather than report an experimentally validated method. We hypothesize that effective intracellular cryoprotection *could* be achieved by generating cryoprotective solutes *in situ* from a latent precursor state, activated by a physical trigger that naturally accompanies freezing under isochoric conditions, the increase in pressure. The purpose of this contribution is to articulate the thermodynamic basis and physical plausibility of this hypothesis and to delineate its potential implications.

Previous approaches in cryobiology have explored nanoparticle-mediated intracellular delivery of pre-formed cryoprotectants such as trehalose, [1–4] in which cryoprotection depends on osmotically active solutes introduced prior to freezing or released through biologically mediated triggers [5], including pH-responsive [6]or temperature-responsive mechanisms [7]. In contrast, the present work considers solute generation driven by the thermodynamic state variables intrinsic to isochoric freezing, pressure, [8], without reliance on biochemical triggers or membrane transport kinetics.

Within the context of the present hypothesis, we consider supramolecular assemblies formed directly from cryoprotectant monomers or oligomers. In their assembled state, these structures are hypothesized to behave as osmotically silent or weakly osmotically active entities relative to the fully dissociated monomers. Typically, such supramolecular assemblies would be *conceptually engineered* to be sufficiently hydrophobic or amphiphilic to permit cellular uptake through passive membrane partitioning and transient bilayer permeation, or alternatively through energy-dependent endocytic pathways, while remaining in a latent, osmotically inactive state prior to pressure-induced disassembly [1–4].

Cryoprotective function, whether in the vasculature or intracellular space, is hypothesized to arise upon pressure activated disassembly, which releases the constituent monomers in free, osmotically active form. In this framework, the supramolecular assembly is treated not as a chemically specific construct, but as an abstract thermodynamic state, introduced to evaluate physical plausibility rather than to define a realizable formulation.

**Helmholtz Free Energy Theory**

Isochoric freezing imposes a strict constant-volume constraint on the system. Under these conditions, the appropriate thermodynamic potential governing equilibrium and stability is the Helmholtz free energy.

$$A = U - TS$$

[1]

whose natural variables are temperature $T$, volume $V$, and composition.

The total differential of the Helmholtz free energy is

$$dA = -S\, dT - P\, dV + \sum_i \mu_i\, dN_i$$

[2]

Under isochoric conditions, $dV = 0$, and therefore changes in the Helmholtz free energy are not directly associated with mechanical work. Pressure does not appear as an independent energetic contribution but arises implicitly through the system's equation of state,

$$P = -\left(\frac{\partial A}{\partial V}\right)_T$$

[3]

Thus, under constant volume, pressure is a *dependent variable* reflecting the curvature of the Helmholtz free-energy surface.

We consider two thermodynamic states of the same closed system:
(i) a supramolecularly assembled state, and
(ii) a dissociated state consisting of free cryoprotectant monomers or oligomers in solution.

The stability of the supramolecular assembly under isochoric conditions is governed by the Helmholtz free-energy difference

$$\Delta A = A_{\text{assembled}} - A_{\text{dissociated}}$$

[4]

At low pressure, supramolecular assembly is favored when $\Delta A < 0$, typically due to favorable enthalpic interactions and modest entropic penalties. Many non-covalent assemblies exhibit a positive excess assembly volume, arising from solvent structuring, excluded volume effects, or packing inefficiencies relative to the dissociated state.

During isochoric freezing, cooling below the equilibrium freezing temperature leads to partial ice formation. Because the system volume is fixed, this phase change produces a substantial increase in hydrostatic pressure. This pressure increase is not an external input but a thermodynamic consequence of changes in the Helmholtz free-energy landscape under constant volume. In the context of this isochoric hypothesis, it is more precise to refer to this as the "pressure-dependence of the chemical equilibrium" or the "volume-pressure contribution to the Helmholtz free energy".

Elevated pressure alters the relative stability of the assembled and dissociated states through their differing molar volumes. For assemblies with positive excess assembly volume, increasing pressure favors the dissociated state, which occupies a smaller effective volume. In Helmholtz-free-energy terms, pressure reshapes the free-energy surface such that the minimum corresponding to the assembled state is destabilized relative to the dissociated state (Fig 1).

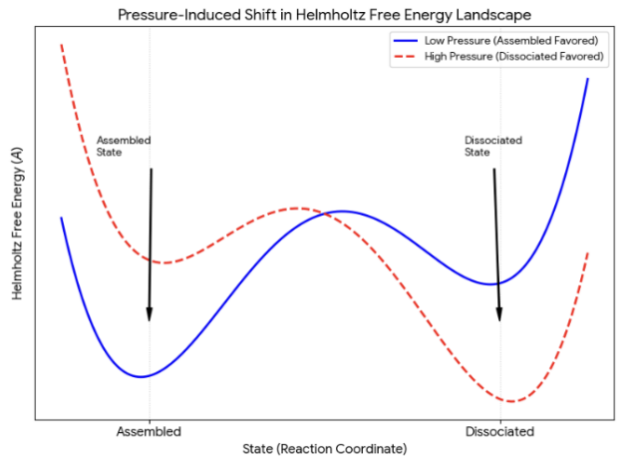

**Figure 1. This diagram illustrates the thermodynamic hypothesis proposed in the paper, showing how the Helmholtz free-energy landscape shifts during isochoric (constant-volume) freezing.**

In the temperature and pressure ranges relevant to isochoric freezing of aqueous soft-matter systems, the pressure dependence of internal energy and entropy is weak. These systems are nearly incompressible, undergo no chemical reactions, and experience only structural rearrangements. Changes in entropy with pressure are modest, and at subzero temperatures the entropic contribution to the free energy is further reduced.

As a result, the dominant factor governing the shift in stability between assembled and dissociated states is the pressure sensitivity of the Helmholtz free energy associated with volume differences between these states. Pressures generated during isochoric freezing, approximately 10 MPa near −1 °C and exceeding 200 MPa near −21 °C, are sufficient to substantially perturb supramolecular equilibria.

Figure 1 illustrates that under normal conditions or the start of the process, the system resides in the "Assembled State" minimum. In this state, the cryoprotectants are in supramolecular assemblies that are osmotically weakly active. This allows them to be loaded into cells or tissues without causing osmotic injury. As cooling-induced ice formation occurs within the fixed volume, hydrostatic pressure increases substantially (up to 200 MPa). Because the dissociated state (monomers) occupies a smaller effective molar volume than the assembled state, the elevated pressure lowers the free energy of the dissociated state relative to the assembly. When the pressure is high enough, the Helmholtz free-energy difference ($\Delta A$) changes sign, rendering the assembled supramolecular state thermodynamically unfavorable and driving spontaneous disassembly into a solution of cryoprotectant monomers or oligomers. The "minimum" on the right becomes the globally stable state, driving the spontaneous disassembly of the particles into active cryoprotective solutes exactly when and where they are needed to prevent ice-related injury. This mechanism effectively uses pressure as a "switch" to activate cryoprotection.

This pressure-activated disassembly mechanism enables controlled generation of high intracellular cryoprotectant concentrations without osmotic loading prior to cooling. Because pressure arises naturally during isochoric freezing, solute generation is intrinsically synchronized with the physical onset of freezing. No changes in pH, ionic strength, or chemical composition are required.

Importantly, the pressure and temperature at which disassembly occurs may be tuned by design. Because the pressure sensitivity scales directly with the volume difference between assembled and dissociated states, supramolecular architecture and ligand chemistry provide control parameters for triggering disassembly at desired thermodynamic conditions.

The framework applies specifically to assemblies whose stability decreases with pressure and assumes that pressure levels generated during isochoric freezing are sufficient to perturb assembly equilibria on relevant timescales. Failure of either condition would falsify the mechanism for a given system.

For single cells, pressure-activated solute generation offers a route to intracellular cryoprotection without membrane transport. In tissues and organs, the framework suggests coordinated intracellular and extracellular cryoprotection driven by a global physical variable rather than by transport kinetics. Crucially, for intracellular cryoprotection the loading of these assemblies must occur at physiological temperatures (e.g., 37°C) to maintain the metabolic activity required for active endocytosis. This 'Warm Loading, Cold Trigger' approach ensures that the cells are fully laden with latent cryoprotectants prior to the onset of isochoric cooling, at which point the rising pressure serves as the autonomous activation switch

The scientific literature provides several experimental parallels that support the thermodynamic pressure-trigger mechanism proposed in this work. Although the specific application to cryopreservation is novel, the behavior of supramolecular assemblies under high hydrostatic pressure (HHP) is well documented. Numerous experimental studies on protein complexes and supramolecular polymers demonstrate that pressures in the range encountered during isochoric freezing (approximately 10–200 MPa) are sufficient to disrupt non-covalent interactions [9–11]. In particular, pressures as low as 10–40 MPa have been shown to induce dissociation of protein quaternary structures.

This pressure-induced dissociation occurs because the separated state, consisting of individual monomers surrounded by hydration shells, often occupies a smaller total volume than the intact assembly, directly supporting the volume-change ($\Delta V$) argument advanced in this paper. Experimental studies have shown that pressures in the 10–40 MPa range can dissociate protein complexes, likely due to tighter packing of water molecules around the dissociated components [12,13]. These studies explicitly revisit the thermodynamic basis for pressure-induced protein unfolding and emphasize the central role of volume changes in such transitions.

A general principle in high-pressure science is that processes which reduce the overall system volume are thermodynamically favored. The system volume is typically lower when individual components are highly hydrated than when they are assembled into a complex in which polar or ionic groups are partially buried. Increasing pressure therefore drives the system toward a lower-volume, disassembled, and more highly hydrated state.

Two closely related phenomena, electrostriction and enhanced hydration, play key roles in this process. Electrostriction refers to the compression and ordering of solvent molecules (primarily water) around charged or polar groups on a molecule, resulting in a hydration shell that is more densely packed than bulk water. Enhanced hydration occurs when a supramolecular assembly, such as a protein-protein complex, dissociates under pressure, allowing water molecules to penetrate the interfacial region and hydrate previously buried polar or ionic groups.

Upon disassembly, formerly "hidden" hydrophobic or charged surfaces become exposed to the solvent. Water molecules pack more tightly around these newly exposed surfaces than in the bulk phase. Consequently, even though the number of individual particles (monomers) increases, the total volume of the system decreases because the surrounding water effectively "contracts" around the hydrated molecules. This reduction in system volume renders the dissociated state thermodynamically favorable at elevated pressures [14] [15]

**Analysis of Helmholtz Free Energy Systems**

To assess the physical plausibility of the proposed hypothesis, we perform an order-of-magnitude thermodynamic analysis to estimate the molar volume change ($\Delta V$) that would be required, within the context of the hypothesis, to induce disassembly of a supramolecular complex at pressures attainable during isochoric freezing.

A supramolecular assembly is stable at atmospheric pressure, $P_{initial}$, ( 0.1 MPa) if its Helmholtz free energy, A, is lower than that of its dissociated monomers ($\Delta A < 0$). For disassembly to be "triggered" by an increase in pressure, the pressure-dependence of the chemical equilibrium or the volume-pressure contribution to the Helmholtz free energy must overcome this initial stabilization energy ($\Delta A_{initial}$). The conditions for the "trigger" pressure, $P_{trigger}$, for disassociation, where $\Delta V_{molar}$ refers to the molar excess volume difference between assembled and dissociated states rather than the macroscopic system volume, are derived from the first-order pressure expansion of the Helmholtz potential, commonly used in high-pressure physics and associated fluid theories.

$$A(P) = A(P_0) + \left(\frac{\partial A}{\partial P}\right)_T (P - P_0) \quad (5)$$

where $\left(\frac{\partial A}{\partial P}\right)_T = V$

which yields:

$$\Delta A(P) \approx \Delta A_{initial} + \Delta V_{molar}(P_{trigger} - P_{initial}) = 0 \quad (6)$$

While the stability of biological assemblies is traditionally modeled using the Gibbs function, our use of the Helmholtz linear expansion, Eq (6), is a rigorous application of the first-order pressure expansion of the Helmholtz potential, A, commonly utilized in high-pressure physics [16] and associating fluid theories. This form is specifically required here due to the isochoric nature of the system, where pressure is the dependent variable.

Assuming that $\Delta V_{molar}$ is weakly pressure-dependent over the range of interest (linear approximation), the molar volume change required to trigger disassembly is

$$\Delta V_{molar} = -\frac{\Delta A_{initial}}{\left(P_{trigger} - P_{initial}\right)} \qquad (7)$$

For supramolecular assemblies that occupy a larger molar volume than their dissociated monomers ($\Delta V_{molar} > 0$), increasing pressure favors disassembly.

Why do pressure-volume terms still appear in the analysis? Although no external mechanical work is performed on the system as a whole, the pressure-volume term in the formulation reflects changes in the chemical potential (i.e., molar free energy) of the constituent molecules. At the level of the entire system, no mechanical work is done because the total volume remains constant $\Delta V_{total} = 0$; pressure is therefore a dependent variable that emerges from the system's internal state.

At the molecular level, however, the situation is different: the molar volume changes ($\Delta V_{molar} \neq 0$). Even under isochoric conditions, the effective volume occupied by a dissociated monomer differs from that occupied by the same monomer when incorporated into a larger macromolecular assembly. As pressure increases, the Helmholtz free-energy landscape is consequently reshaped. States with smaller molar volumes, such as the dissociated state, become energetically more favorable than states with larger molar volumes. Thus, just as heat drives endothermic reactions, pressure under isochoric conditions drives reactions associated with a negative change in molar volume.

Following is a sample calculation. In an isochoric freezing system held at −1.5 °C, the internal pressure reaches approximately 20 MPa. Taking 20 MPa as a representative pressure for subzero isochoric conditions, and considering typical stabilization energies for non-covalent assemblies (on the order of 10–60 kJ mol⁻¹), we estimate the molar volume changes required for pressure-activated decomposition of a supramolecular assembly.

While supramolecules are treated here as an abstract state, physical realizability necessitates that these assemblies do not exceed approximately 40 nm to 50 nm in diameter. This size constraint ensures that the particles are small enough to be internalized by cells via endocytic pathways, yet large enough to sequester thousands of cryoprotectant monomers in an osmotically silent state until the pressure trigger is activated. Given an optimal 40 nm to 50 nm diameter, significant intracellular concentrations can typically be reached within a 2 to 4-hour incubation window at 37°C. For the application suggested in this hypothetical study, the time for incubation is a critical parameter. The selection of a 40 nm target diameter for the supramolecular assembly is based on established nanomedicine benchmarks, which identify the 40–50 nm range as the peak efficiency for clathrin-mediated endocytosis due to the optimal balance between receptor recruitment and membrane bending energy [17,18]. The size of nanoparticles is a parameter shown by Bischof and colleagues to be critical for modulating thermal and phase-change properties during cryobiological procedures [19]. The rate of supramolecular incorporation is a non-linear function of diameter, peaking at about 40-50 nm. Deviating to a diameter of 10 nm or 100 nm can result in a 60–80% reduction in uptake velocity compared to the 40 nm optimum, primarily due to the energetic costs of membrane wrapping and vesicle scission.

Basic volume calculations show that a design for a 40 nm to 50 nm supramolecule size would incorporate between 2000 and 3000 molecules of glycerol, about 800 molecules of trehalose or about 1000 molecules of glucose. Therefore loading the cell with a 0.01 mol of the supramolecule could result in vitrification levels of cryoprotectants inside the cell, once activated by isochoric freezing.

The analysis assumes the existence of supramolecular assemblies comprising on the order of $10^3$ individual cryoprotectant molecules, such as glycerol, glucose, or trehalose, organized into a single pressure-responsive entity. At this aggregation level, the assembled state is effectively osmotically silent or weakly active, while disassembly yields a large number of freely dissolved solute molecules and a correspondingly large increase in osmotic activity. The purpose of the supramolecular architecture in this context is not to prescribe a specific chemistry, but to realize a physically plausible difference in molar volume between the assembled and dissociated states, $\Delta V_{rxn} = V_{ass} - V_{dis}$, that can couple internal chemical transformations to the pressure field generated during isochoric freezing.

A variety of supramolecular design strategies could, in principle, be used to achieve such pressure-sensitive architectures. One illustrative route involves amphiphilic derivatives of cryoprotectants, in which hydrophilic polyol headgroups are linked to hydrophobic moieties, enabling self-assembly into vesicular or micellar structures. For example, glycerol-derived amphiphiles may be envisaged by transient masking of hydroxyl groups and attachment of hydrophobic chains, leading to assemblies that resemble lipid vesicles but possess a larger effective molar volume due to incomplete packing and internal free volume. When combined with bulky hydrophobic spacers, such as squalene, α-tocopherol, or related sterically demanding molecules, the assembled state can be further expanded, increasing $V_{ass}$ relative to the dissociated solute state. From a thermodynamic perspective, such spacers act as volumetric diluents of the assembly, amplifying the pressure sensitivity through an increased $\Delta V_{rxn}$.

Beyond vesicular architectures, other classes of supramolecular constructs may serve the same thermodynamic function. These include polymeric or oligomeric micelles formed from cryoprotectant-grafted backbones, loosely crosslinked nanogels with high solvent content, dendritic or hyperbranched assemblies with internal cavities, and host-guest complexes stabilized by weak noncovalent interactions. In each case, the common design principle is that the assembled state occupies a larger effective molar volume than the ensemble of dissociated cryoprotectant molecules in solution, such that increasing pressure shifts the Helmholtz free-energy balance in favor of disassembly. Importantly, the precise chemical identity of the assembly is secondary to this volumetric criterion; the assemblies need only be metastable under ambient conditions and destabilized when the pressure term $P \Delta V_{rxn}$ becomes comparable to or exceeds the intrinsic stabilization free energy of the supramolecular state.

Within the Helmholtz framework adopted here, these supramolecular constructs function as latent solute reservoirs whose activation is governed by pressure rather than temperature or membrane transport. As pressure rises during isochoric freezing, the $P \Delta V_{rxn}$ contribution to the free energy lowers the stability of the assembled state, triggering disassembly and release of cryoprotectant molecules precisely at the onset of ice formation. The resulting increase in solute concentration reshapes the liquid Helmholtz surface and modifies the equilibrium partition between ice and liquid. While the present work does not propose a specific synthetic implementation, it establishes the thermodynamic criteria that any such pressure-activated supramolecular system must satisfy, thereby defining a design space rather than a recipe.

By utilizing a 40 nm polycarbonate membrane during extrusion, the resulting particles can be precisely sized to maximize uptake via clathrin-mediated endocytosis, a process that is biologically optimal in the 20–50 nm range. The inclusion of these spacers is critical to the isochoric hypothesis, as they create the internal 'void volumes' required to satisfy the condition for pressure-activated disassembly.

The use of stabilization energies in the range of 10–50 kJ mol$^{-1}$ is drawn from well-established reference values in supramolecular chemistry and molecular biology for non-covalent interactions. This range represents a "stability window" in which assemblies are sufficiently stable to persist in biological environments, yet weak enough to remain reversible or responsive to an external trigger, such as the pressure generated during isochoric freezing [20] [21]. In computational chemistry, the so-called "gold standard" for binding energies of molecular complexes typically targets a similar range, approximately 10–60 kJ mol$^{-1}$ [22]. Energies below this range correspond to transient, non-specific interactions, whereas significantly higher values approach the strength of weak covalent bonds or strong metal–ligand coordination.

This energetic window is therefore ideally suited for a pressure-activated switching mechanism: assemblies remain stable during storage and loading, yet become unstable under the specific physical stress imposed by isochoric cooling. Notably, the non-covalent binding energy range associated with "chemical accuracy" (approximately 10–60 kJ mol$^{-1}$) aligns well with the pressure levels (≈10–200 MPa) attainable during isochoric freezing, providing a natural thermodynamic match between molecular stability and the pressure scales involved [22].

Table 1 summarizes the total and per-monomer volume changes required to overcome different initial stabilization energies for a supramolecular assembly with a representative size of monomers. Calculations are provided for pressures typical of subzero isochoric freezing (20 MPa) and higher pressures (200 MPa).

Table 1. Estimated molar volume changes for pressure-activated disassembly (N=1000).

| Initial Stability | Trigger Pressure | Required (Total) | Required per monomer |
| --- | --- | --- | --- |
| 10 kJ/mol (Weak) | 20 MPa | ~500 cm³/mol | ~0.50 cm³/mol |
| 30 kJ/mol (Moderate) | 20 MPa | ~1500 cm³/mol | ~1.51 cm³/mol |
| 50 kJ/mol (Strong) | 20 MPa | ~2500 cm³/mol | ~2.51 cm³/mol |
| 30 kJ/mol (Moderate) | 200 MPa | ~150 cm³/mol | ~0.15 cm³/mol |

Within the context of the proposed hypothesis, using an assembly size to 1000 monomers significantly enhances the thermodynamic plausibility of the system. For a representative pressure of 20 MPa, the required total volume reductions is between 500–2500 cm³ mol⁻¹. However, when distributed across 1000 monomers, the required per-monomer volume changes drop to a range of roughly 0.50–2.51 cm³ mol⁻¹.

These changes are remarkably small relative to the molar volumes of common cryoprotectants. For glycerol ($\approx$ 73 cm³ mol⁻¹), a change of 0.5–1.5 cm³ mol⁻¹ represents a difference of only 0.7–2.0% in effective volume. For glucose ($\approx$ 112 cm³ mol⁻¹), a required change of 0.5 cm³ mol⁻¹ represents less than 0.5% of its molar volume. For trehalose ($\approx$ 210 cm³ mol⁻¹), at 200 MPa, the required per-monomer volume change of ~0.15 cm³ mol⁻¹ is less than 0.1% of the molecular volume.

These extremely low requirements suggest that even very stable supramolecular structures could be triggered to disassemble under minimal changes in packing efficiency or hydration states during isochoric freezing.

The table also illustrates the strong sensitivity of the pressure-activation mechanism to pressure magnitude. Increasing the pressure from 20 MPa to 200 MPa reduces the required total volume change by an order of magnitude for the same stabilization energy (30 kJ mol⁻¹), lowering the per-monomer requirement to values that are extremely modest. Such pressures are achievable in isochoric systems near −21 °C, underscoring the feasibility of using isochoric pressure as a controllable thermodynamic trigger.

Overall, Table 1 is consistent with the hypothesis advanced in this work: that modest, physically reasonable molar volume differences between assembled and dissociated states could, in principle, permit pressure-activated disassembly of supramolecular cryoprotectant assemblies under isochoric freezing conditions. The analysis is intended to establish physical feasibility rather than experimental demonstration, and to delineate a set of quantitatively testable predictions. Notably, the inferred per-monomer volume changes ($\approx 0.5 - 2.5$ cm³ mol⁻¹) are comparable in magnitude to experimentally reported hydration- and pressure-induced volume changes associated with protein unfolding and carbohydrate solvation, indicating that the proposed mechanism does not rely on anomalously large molecular rearrangements.

From a design perspective, the pressure at which a supramolecular assembly undergoes disassembly is expected to be tunable through general architectural and hydration-related features. Assemblies that incorporate internal voids, loose packing, or significant solvent-filled regions are anticipated to be more sensitive to pressure, as these features increase the potential volume reduction upon disassembly. Likewise, assemblies with inefficient internal packing, should exhibit a stronger thermodynamic drive toward dissociation under elevated pressure, since the dissociated, hydrated state occupies a smaller effective volume. The pressure threshold for activation may also be influenced by the hydration characteristics of the monomers, particularly when disassembly exposes polar or hydrophobic surfaces that become densely hydrated. Together, these considerations suggest that pressure-activated disassembly during isochoric freezing is not a fixed material property, but a tunable response that can be guided by established principles of supramolecular chemistry and high-pressure science.

**Gibbs Free Energy Theory**

While isochoric freezing provides a convenient and intrinsic mechanism for generating elevated pressure, the underlying concept is not limited to isochoric systems. In principle, analogous effects can be achieved by externally applied mechanical pressure under isothermal or non-isochoric conditions. Supramolecular association and dissociation under pressure may be described equivalently using the Helmholtz free energy in isochoric systems or the Gibbs free energy in isobaric systems. In the isochoric formulation, pressure emerges as a dependent variable that reshapes the Helmholtz free-energy landscape, whereas in the isobaric formulation pressure enters explicitly through the Gibbs free energy. In both cases, assemblies with a positive excess molar volume are destabilized by increasing pressure, and dissociation occurs when the pressure-dependent free-energy difference between assembled and dissociated states changes sign. The thermodynamic driving force for disassembly is therefore independent of the mechanism by which pressure is generated, although isochoric freezing offers a natural and synchronized means of achieving the required pressure levels during cooling.

For an isobaric system in which pressure is externally applied and increased mechanically from ambient conditions ($P_0 \approx 0.1$ MPa) to pressures comparable to those generated during isochoric freezing, the appropriate thermodynamic potential governing equilibrium is the Gibbs free energy, G,:

$$G = U + PV - TS \qquad (8)$$

with variables: internal energy, U, entropy, S, temperature, T, Pressure, P, Volume, V, and composition ($\{N_i\}$).

The total differential of the Gibbs free energy is:

$$dG = -S\, dT + V\, dP + \sum_i \mu_i\, dN_i \qquad (9)$$

At constant temperature and composition:

$$\left(\frac{\partial G}{\partial P}\right)_T = V \qquad (10)$$

Thus, the pressure dependence of the Gibbs free energy is governed directly by the system volume.

Similarly to the analysis in an isochoric system, consider two thermodynamic states of the same closed system:

**A**: assembled supramolecular structure and **D**: dissociated monomers or oligomers in solution

We define the Gibbs free-energy difference between the assembled and dissociated state as:

$$\Delta G(P) \equiv G_A(P) - G_D(P) \qquad (11)$$

and the corresponding volume difference:

$$\Delta V \equiv V_A - V_D \qquad (12)$$

From the differential relation, we obtain:

$$\left(\frac{\partial \Delta G}{\partial P}\right)_T = \Delta V \qquad (13)$$

Integrating from ambient pressure $P_0$ to an elevated pressure $P$:

$$\Delta G(P) = \Delta G(P_0) + \int_{P_0}^{P} \Delta V(P')\, dP' \qquad (14)$$

For supramolecular assemblies in aqueous solution, compressibility is low and $\Delta V$ varies weakly with pressure over the range of interest. Under this approximation:

The pressure dependence of the free energy may be written, to first order, as

$$\Delta G(P) \approx \Delta G(P_0) + \Delta V (P - P_0) \qquad (15)$$

which is the central result used here. This linearized expression is valid provided that the excess volume change $\Delta V$ is approximately pressure-independent over the range of interest, an assumption commonly employed in high-pressure thermodynamics.

When the supramolecular assembly has a positive excess volume ($\Delta V > 0$), a situation typical of many assemblies, the assembled state occupies a larger effective volume than the dissociated state. This positive excess volume commonly arises from solvent exclusion, restructuring of hydration shells, and packing inefficiencies within the assembled structure. Under these conditions, the free-energy difference $\Delta G(P)$ increases linearly with pressure. At sufficiently high pressure, a critical pressure $P^*$ is reached at which the free-energy difference vanishes, $\Delta G(P^*) = 0$. Solving for this condition yields

$$P^* = P_0 - \frac{\Delta G(P_0)}{\Delta V} \qquad (16)$$

For pressures exceeding $P^*$, dissociation becomes thermodynamically favored.

In contrast, when the assembly has a negative excess volume ($\Delta V < 0$), the assembled state becomes increasingly stable as pressure rises. Such systems are therefore not expected to dissociate under either isochoric pressure generation or externally applied mechanical pressure.

The pressure dependence of the assembly equilibrium may also be expressed in terms of the equilibrium constant $K(P)$, defined by

$$K(P) = \exp\left[-\frac{\Delta G(P)}{RT}\right] \qquad (17)$$

Substituting the linearized pressure dependence of the free energy yields

$$\ln\left[\frac{K(P)}{K(P_0)}\right] = -\frac{\Delta V}{RT}(P - P_0) \quad (18)$$

This relation predicts an exponential suppression of assembly with increasing pressure when $\Delta V > 0$, enables continuous tuning of the equilibrium constant by pressure, and provides a direct route for experimental validation.

**Comparison of isochoric and isobaric pressure triggers**

In isochoric freezing, pressure arises as a dependent variable imposed by the thermodynamic constraint of constant volume during cooling, whereas in an isobaric formulation pressure is externally applied. In both cases, the effect on supramolecular association–dissociation is governed solely by the pressure dependence of the free energy, $\Delta G(P)$, and is therefore independent of the mechanism by which pressure is generated. Isochoric freezing thus provides a natural and synchronized means of achieving pressure-activated disassembly during cooling, although the underlying thermodynamic mechanism is general.

In an isobaric system at constant temperature, the pressure dependence of supramolecular assembly is governed by the Gibbs free energy. The free-energy difference between assembled and dissociated states varies with pressure according to $\Delta G(P) = \Delta G(P_0) + \Delta V(P - P_0)$, where $\Delta V$ is the excess volume of the assembled state. For assemblies with positive excess volume, increasing pressure destabilizes the assembled state and favors dissociation. Pressure increases comparable to those generated during isochoric freezing are therefore sufficient to reverse supramolecular assembly equilibria even when applied mechanically under isobaric conditions.

While pressure-activated disassembly of supramolecular assemblies can, in principle, be achieved by mechanically imposing pressure under isobaric conditions, isochoric freezing offers several fundamental advantages. In an isochoric system, pressure is generated intrinsically as a thermodynamic consequence of cooling and phase evolution at fixed volume, eliminating the need for external pressure sources, seals, or mechanical actuation. The resulting pressure increase is spatially uniform, temporally synchronized with the onset and progression of freezing, and inherently coupled to the thermodynamic state of the system, ensuring that disassembly occurs precisely when cryoprotection is required. In contrast, mechanically applied isobaric pressurization requires independent control, introduces engineering complexity, and risks temporal or spatial mismatch between pressure application and the freezing process. Moreover, pressure generation under isochoric conditions is self-limiting and fully reversible upon warming, reducing the likelihood of overexposure to damaging pressure levels.

An additional and important advantage of pressure generation by isochoric freezing is its intrinsic biological safety. Hydrostatic pressure is known to exert detrimental biological effects, particularly at elevated or near-physiological temperatures, such that mechanical pressurization under isobaric conditions requires careful temporal separation from biologically sensitive temperature regimes. In contrast, isochoric freezing generates pressure only as ice forms, precisely when reduced molecular mobility and low temperature mitigate pressure-induced damage. Pressure exposure is therefore naturally confined to the cryogenically relevant temperature range. Furthermore, because pressure under isochoric conditions increases continuously with the extent of ice formation, pressure-activated disassembly can occur progressively during cooling. By designing supramolecular assemblies with tailored activation pressures, this mechanism enables a gradual, self-regulated increase in intracellular cryoprotectant concentration as freezing proceeds, rather than a single abrupt release. This continuous coupling between freezing progression, pressure elevation, and solute generation provides a degree of thermodynamic control that is difficult to achieve with externally imposed isobaric pressurization.

**Conclusion**

This Communication advances a *thermodynamic hypothesis*, formulated within a Helmholtz free-energy framework, proposing that cryoprotective solutes *could* be generated in situ through pressure-activated disassembly of supramolecular assemblies during isochoric freezing. The hypothesis presentation is intentionally conceptual and feasibility-oriented, and is not intended to demonstrate or verify such behavior experimentally. By decoupling membrane transport from cryoprotection and treating pressure as an intrinsic thermodynamic control variable rather than an externally applied work input, the proposed hypothesis delineates a new mechanistic pathway that *may* be explored experimentally. As such, the framework is offered as a basis for testable predictions and future investigation, rather than as an established cryoprotective mechanism.

**Funding:** This research received no external funding

**Conflict of Interests:** The author has a minority finnancial stake in BioChoric a commercial entity in the field of isochoric technologies.

**Acknowledgement:** The authors used ChatGPT and Gemini to assist with the linguistic refinement of the manuscript. The final version was reviewed and approved by the author